\documentclass[
 reprint,
 amsmath,
 pre,
 amssymb,
 nofootinbib,
 aps,
]{revtex4-2}

\usepackage[pdftex]{graphicx}
\usepackage{wasysym}
\usepackage{amsfonts}
\usepackage{ifsym}
\usepackage{braket}

\usepackage{amsmath,amssymb,amsfonts,amscd,mathrsfs}
\usepackage{dsfont}

\usepackage{lipsum}
\usepackage{pifont}

\usepackage{tikz-feynman}
 \tikzfeynmanset{compat=1.0.0}

\makeatletter
\newlength{\apb@width}
\newcommand{\autoparbox}[2][c]{\settowidth{\apb@width}{#2}\parbox[#1]{\apb@width}{#2}}

\makeatother

\newcommand{\namedref}[2]{\hyperref[#2]{#1~\ref*{#2}}}

\newcommand{\be}{\begin{equation}}
\newcommand{\ee}{\end{equation}}

\newcommand{\reef}[1]{(\ref{#1})}

\def\bea#1\eea{\begin{align}#1\end{align}}

\newcommand{\eps}{\varepsilon}

\newcommand{\Csphere}{{}^\bullet\kern-1.2pt C}
\newcommand{\Ctorus}{{}^\circ\kern-1.2pt C}

\newcommand{\COMMENT}[1]{}

\newcommand{\neqa}{\nonumber\end{eqnarray}}

\newcommand{\<}{{\langle}}
\renewcommand{\>}{{\rangle}}

\newcommand{\re}{\relax{\rm I\kern-.18em R}}

\def\su2{{SU(2)}}

\def\eps{{\epsilon}}

\def\[{\left[}
\def\]{\right]}

\def\({\left(}
\def\){\right)}
\def\[{\left[}
\def\]{\right]}

\def\<{\langle}
\def\>{\rangle}

\def\i2{\frac{i}{2}}

\def\2F1{\,_2{\rm F}_1}

\usepackage{color}
\usepackage{graphicx}
\usepackage{dcolumn}
\usepackage{bm}
\usepackage{hyperref}

\usepackage{cancel}
\usepackage{ctable}
\usepackage{booktabs}
\newcolumntype{L}[1]{>{\raggedright\let\newline\\\arraybackslash\hspace{0pt}}m{#1}}
\newcolumntype{C}[1]{>{\centering\let\newline\\\arraybackslash\hspace{0pt}}m{#1}}
\newcolumntype{R}[1]{>{\raggedleft\let\newline\\\arraybackslash\hspace{0pt}}m{#1}}

\newcommand{\beq}{\begin{equation}}
\newcommand{\eeq}{\end{equation}}
\newcommand{\beqq}{\begin{equation*}}
\newcommand{\eeqq}{\end{equation*}}
\newcommand\beqa{\begin{eqnarray}}
\newcommand\eeqa{\end{eqnarray}}
\newcommand\beqaa{\begin{eqnarray*}}
\newcommand\eeqaa{\end{eqnarray*}}

\begin{document}

\title{ 
Flowing from the Ising Model on the Fuzzy Sphere to the 3D Lee-Yang CFT
}

\author{Joan Elias Mir\'o$^{~ (a)}$}
\affiliation{\vspace{.2cm}  (a) The Abdus Salam ICTP,    Strada Costiera 11, 34135, Trieste, Italy }
\author{Olivier Delouche$^{~ (a,b)}$}
\affiliation{\vspace{.2cm}   (a) The Abdus Salam ICTP,    Strada Costiera 11, 34135, Trieste, Italy }
\author{~}
\affiliation{
\vspace{-.7cm} (b) Université de Genève,  24 quai Ernest-Ansermet, 1211 Genève 4, Switzerland
}

\begin{abstract}

We employ the Fuzzy Sphere regulator to study the 3D Lee-Yang CFT. The model is defined by deforming the Ising model on the Fuzzy Sphere via a purely imaginary longitudinal magnetic field. This model undergoes a quantum phase transition, whose critical point we determine and identify with the 3D Lee-Yang CFT. We show how to tune the model and find that the lowest-lying states of the Hamiltonian align well with the expected CFT spectrum. We discuss the Fuzzy Sphere estimates for the scaling dimension $\Delta_\phi$ of the lowest primary operator.
Finally, we interpret small  deviations from the CFT expectations in terms of the leading irrelevant operators of the Lee-Yang CFT. 
We show that the Fuzzy Sphere calculations are compatible with the best  five-loop $\epsilon$-expansion estimates. 

\end{abstract}

\pacs{Valid PACS appear here} 

\maketitle


\section{Introduction}

In this work, we build on the recent development of the 3D Ising model on the fuzzy sphere~\cite{Zhu:2022gjc} and study the renormalisation group (RG) flow triggered by  the ``$i \sigma$" deformation, which flows to the Lee–Yang (LY) CFT.

The origin of the LY CFT and its role in our understanding of order–disorder phase transitions is a remarkable chapter in theoretical physics; see Ref.~\cite{Cardy:2023lha} for a recent review and historical remarks.
The universal, long-distance physics,  of the LY edge singularity~\cite{Lee:1952ig,Yang:1952be} is captured  by the following field theory action~\cite{Fisher:1978pf}
\be
S_\text{LY}= \int d^{6-\eps}x \left( \frac{1}{2}(\partial\phi)^2+ \frac{i \lambda}{3!}\phi^3\right) \, . \label{GL}
\ee
The coupling $i\lambda$ is purely imaginary, rendering the theory non-unitary while preserving stability by avoiding runaway directions.~\footnote{The purely imaginary coupling implies the theory has a pseudo-Hermitian, not Hermitian, Hamiltonian—details follow below.}

The free scaling dimension of $\phi^3$ is $3(D - 2)/2$. This operator is classically marginal in $D = 6$ spacetime dimensions, allowing for a Wilson–Fisher~\cite{Wilson:1971dc}  $\epsilon$-expansion to search for a fixed point and compute the anomalous dimensions of the lowest-lying operators.

The current state-of-the-art five-loop $\epsilon$-expansion estimates yield $\Delta_\phi = 0.215(1)$ for $D = 3$~\cite{Borinsky:2021jdb} -- this value lies below the unitarity bound.~\footnote{
In 2D the LY CFT is the $M(2,5)$ minimal model, $\Delta_\phi = -2/5$.~\cite{Cardy:1985yy}}
The next operator in the spectrum is the spin $\ell{=}1$ descendant $\partial_\mu \phi$, with dimension $\Delta_\phi{+}1$.
The operator $\phi^2$ is redundant with $\Box\phi$ via the equations of motion $\Box\phi \propto\phi^2$.
There are only  two level-two descendants $\Box \phi$ ($\ell{=}0$) and the trace-less symmetric $\partial_\mu \partial_\nu \phi$ ($\ell{=}2$). 
Next, as in any local  CFT, the LY has an energy momentum tensor $T^{\mu\nu}$: a spin-two primary operator of dimension $\Delta_T{=}3$. 
The next primary operators in the spectrum are  a scalar operator  $\phi^3$ and a spin $\ell{=}4$ operator $C_{\mu\nu\rho\sigma}$, which have  estimated  dimensions above  three -- indeed $\Delta_{\phi^3}{=}5.0(1)$ and $\Delta_{C}{=} 4.75(1)$~\cite{Gliozzi:2014jsa}. 

Accordingly, the spectrum of low-lying operators is expected to be ordered by ascending scaling dimension as
$[
\mathds{1},\, \phi,\, \partial \phi,\, \partial \partial \phi {\sim} \Box \phi,\, T,\, \partial \Box \phi {\sim} \partial \partial \partial \phi,\, \dots
]$ where we indicate degenerate operators as $\mathcal{O}_1\sim \mathcal{O}_2$.
Their   dimensions are $\[0,\Delta_\phi,\Delta_\phi{+}1,\Delta_\phi{+}2,\Delta_\phi{+}2,3, \Delta_\phi{+}3, \Delta_\phi{+}3 , \dots  \]$, respectively. 

One of the main results of this work is a precise confirmation of these expectations. We also verify the expected conformal spectrum of irrelevant descendants up to $\Delta \leq 5.2$, as well as provide estimates for the scaling dimensions of the irrelevant primaries $\phi^3$ and $C_{\mu\nu\rho\sigma}$. This is achieved through a detailed explanation of how to tune to the LY CFT and the use of Conformal Perturbation Theory (CPT), thereby setting the stage for further measurements of its parameters.
 
\section{A Fuzzy Sphere model for the Lee-Yang CFT}

 The LY CFT  admits a realisation as a quantum critical point. This realisation will be the primary focus of the present work.~\footnote{Two-dimensional studies of this phase transition include a real-space RG analysis~\cite{Uzelac:1979vf} and exact diagonalization of the Ising spin chain in a purely imaginary field~\cite{vonGehlen:1991zlm}, the latter being conceptually closest to our work, albeit in lower dimensions. This 2D RG flow can also be studied via TCSA/TFFSA (a.k.a. Hamiltonian Truncation), see ~\cite{Xu:2023nke,Xu:2022mmw} and references therein. }
We will study a Hamiltonian $H$ defined by deforming the off-critical Fuzzy Sphere Ising Hamiltonian $H_\text{FSI}$ on the Fuzzy Sphere. Let us now briefly introduce the Fuzzy Sphere framework within which this Hamiltonian is built.

The Fuzzy Sphere formulation~\cite{Zhu:2022gjc}  introduces non-relativistic  spin-$\tfrac{1}{2}$ electrons on a  sphere with a $ 4\pi s$ magnetic monopole at the centre. 
The model is restricted to the lowest Landau level (LLL), which are the lowest energy states. 
This  implies that each electron has  $2s + 1$ orbitals (for each spin $\sigma=\uparrow,\downarrow$) in which the electrons can be spanned $\psi_i(\Omega) = \frac{1}{\sqrt{2s+1}}\sum_{m=-s}^s\Phi_m(\Omega) c_{\sigma, m} $, with $\Omega=(\theta, \phi)$ spherical coordinates and $\Phi_m$ monopole harmonics.
The parameter $N$ sets the resolution of the system through the volume of the 2-sphere, ($N$ is proportional to the radius of the sphere squared).
Only half-filled states are considered, in which case the number of electrons   is   $N = 2s + 1$. 

The construction then proceeds by introducing density operators
 $n^a(\Omega) = \psi^\dagger \sigma^a \psi$, where $\sigma^a$ is either the $2\times 2$ identity ($a=0$) or a Pauli matrix ($a=x,y,z$), 
 and defining the Hamiltonian
 \bea
 H_\text{FSI} &= \int (2s+1)^2 d\Omega_{i}d\Omega_{i} U(\Omega_{ij}) \left[n^0_in^0_j- n^z_i n^z_j \right] \nonumber \\
 &- h  \int (2s+1) d\Omega \,  n^x (\Omega)\,,
 \eea
 where $n_i^a\equiv n^a(\Omega_i)$. The potential $U$ introduces local density-density interactions,
 and depends on a finite number of parameters  via a short-distance expansion.
 Following \cite{Zhu:2022gjc} we will only retain the leading two terms, in which case, after fixing dimensions,  it depends on a single real-parameter $V_0$.  
 The second term of the Hamiltonian can be simplified into
 \be
 H_a = \sum_{m=-s}^s {\bf c }^\dagger_m \sigma^a  {\bf c }_m\, ,
  \ee
where  $a=x$, and   ${\bf c}_m^\dagger=(c_\uparrow^\dagger, c_\downarrow^\dagger)$ are fermion creation operators in the $m$th Landau orbital.  
 We refer to \cite{Zhu:2022gjc} for details on  the   Ising Hamiltonian  on the Fuzzy Sphere $H_\text{FSI}$ (there called $H$) -- see also \cite{Fardelli:2024qla} for a lightning review. 

A key feature of this finite-dimensional model is its exact preservation of spatial rotational $SO(3)$ symmetry. As a result, even at finite $N$, the Hamiltonian spectrum organizes neatly into spin-$\ell$ irreducible representations (irreps) of $SO(3)$.

\medskip 

In this work are going to study the following Hamiltonian:
\be
\alpha^{-1} H(V_0,h,g) = H_\text{FSI}(V_0,h) +  i g H_z \, .  \label{LYham}
\ee
The Hamiltonian depends on three real parameters $(V_0, h, g)$ and a normalization constant $\alpha$.

\medskip

The operator $H_z$ breaks the Ising $\mathds{Z}_2$-symmetry. In the Ginzburg-Landau (GL) description \reef{GL} this corresponds to the breaking of the $\phi \to -\phi$ symmetry by the $\phi^3$.

The Hamiltonian of the theory is not Hermitian but satisfies $[H,S]=0$ for an anti-linear symmetry operator $S$. This ensures that the eigenvalues of $H$ are either real or come in complex conjugate pairs. In our case, $S$  acts as $S: H_z \rightarrow -H_z, i \to -i$, and can be associated to a ${\cal PT}$ transformation~\cite{Bender:2018pbv}.

Note that for a finite range of $|g|$, the spectrum of the theory remains real, as explained in~\cite{vonGehlen:1991zlm}. 
At $g = 0$, the lowest energy states of the Ising model (in the disordered phase) are non-degenerate. Thus a finite, but small enough, increment in $g$ cannot lead to complex conjugate pairs. In contrast, for $h = 0$, the vacuum is degenerate, and even an infinitesimal $g > 0$ results in a complex spectrum.

Increasing $g$ is expected to induce a quantum phase transition at the point where eigenvalues merge. At finite $N$, this first occurs for the lowest-lying  states, corres\-ponding to the spin-0 operators  $\mathds{1}$ and $\phi$. For any fixed $(V_0, h)$, we define the critical point $g_c$ as the value at which these two states first become degenerate.
As just explained we expect $g_c$ to vanish as $h\rightarrow 0$. 

\medskip

The Hamiltonian~\reef{LYham} can be seen as a 3D Fuzzy Sphere generalization of the  spin-lattice Hamiltonian studied by von~Gehlen~\cite{vonGehlen:1991zlm}.
This work offers valuable intuition for the present study -- in Appendix~\ref{vG}, we provide a  \textit{Mathematica} code to reproduce it.

\subsection{Critical curve}

 \begin{figure}[t]
  \centering
  \includegraphics[width=0.47\textwidth]{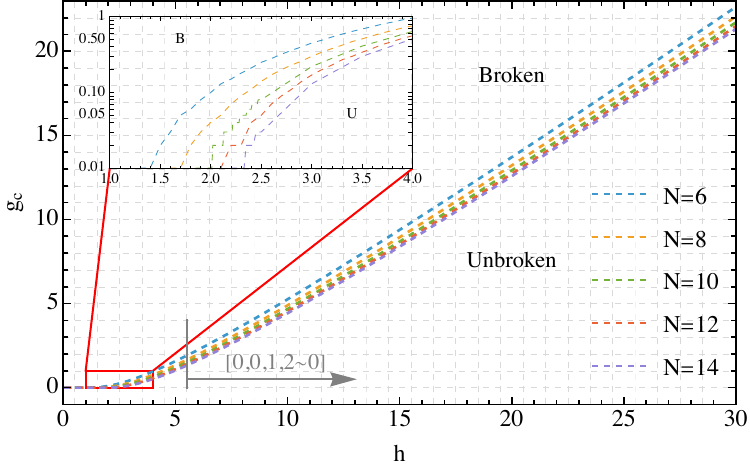}
  \caption{Phase diagram of 3D LY theory. Inset: log-scale zoom.}
  \label{fig:ps}
\end{figure}

\begin{figure*}[t!]
    \centering
    \includegraphics[width=0.33\linewidth]{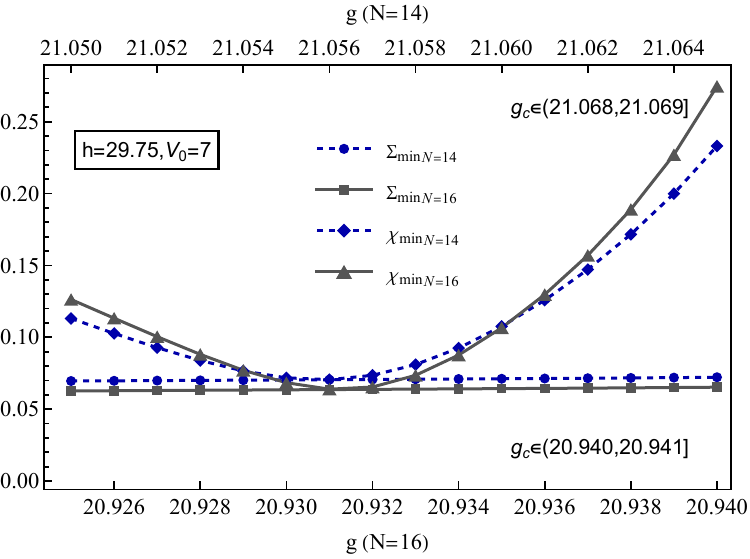}\hfill
    \includegraphics[width=0.335\linewidth]{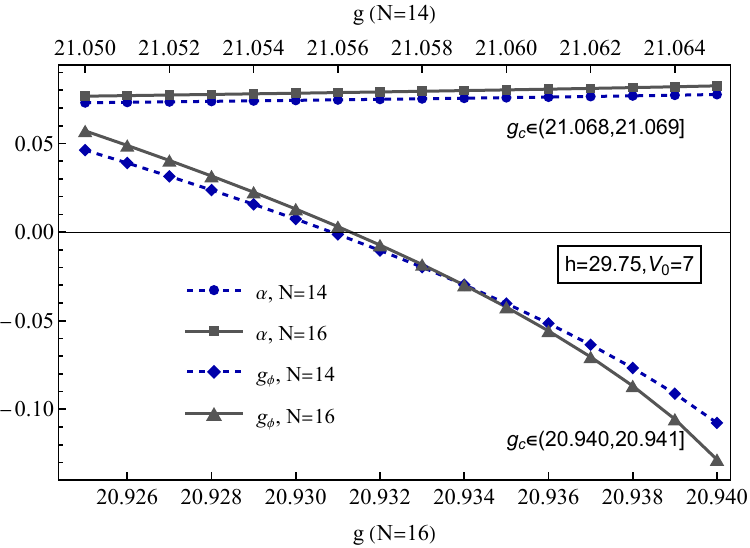}\hfill
    \includegraphics[width=0.32\linewidth]{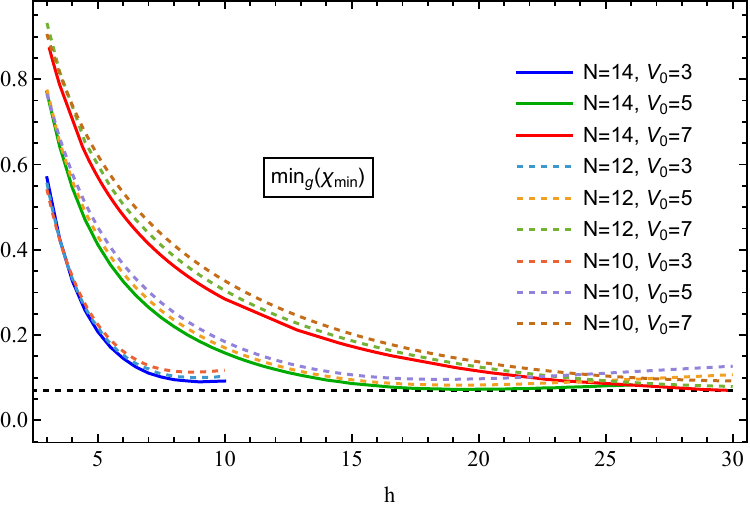}
    \caption{ From left to right: $\chi_\text{min}$ and $\Sigma_\text{min}$  as a function of $g$ just below $g_c$ for $N=14,16$; the  coupling $g_\phi$ minimizing $\Sigma_\text{min}$ and the corresponding $\alpha$ as functions of $g$ for $N=14,16$;  $\text{min}_g\{\chi_\text{min}(g)\}$  as a function of $h$ for $N=10,12,14$ and $V_0=3,5,7$. }
    \label{fig:three_plots}
\end{figure*}
 
 In Fig.~\ref{fig:ps} we show the critical line  $g_c(h)$ for the values $N=6,\, 8,\, \dots, 14$, and $V_0=5$.
 The line is also a function of $N$,  converging monotonically from above with increasing $N$. We suppress the $N$ dependence in `$g_c(h)$' -- which value of $N$ we are referring to will be clear from the context. 
 
For the numerical results, we used our own  \textit{Mathematica} code\footnote{Available upon request.} for $N \leq 8$, while for $N \leq 16$ we used the \texttt{Julia} package \texttt{FuzzifiED}~\cite{Zhou:2025liv}.

 The critical  line is obtained by ``shooting upward'': increasing $g$ by small steps from a value below $g_c(h)$ until the spin-0 operators $\mathds{1}$ and $\phi$ merge into complex conjugates.~\footnote{In practice, since the critical line has positive slope, we find $g_c(h)$ by shooting upward from $g_c(h')$ with $h' < h$.}
For Fig.~\ref{fig:ps}  we  took increasing steps of $\delta g=0.01$ -- the optimal size of the step is addressed below.
The shape of  $g_c(h)$ is independent of   the Hamiltonian normalisation, we thus set $\alpha=1$  in Fig.~\ref{fig:ps}.
 
 The critical surface $g_c(h, V_0)$ depends only weakly on $ V_0$, but the ordering of irreps and the gap values do vary with $V_0$ --  details follow below.

The first interesting check is the spectrum of spin-$\ell$ SO(3) Lorentz irreducible representations (irreps) along the critical line.
The lowest-lying states correspond to the relevant operators ${\mathds{1}, \phi, \partial_\mu \phi, \partial_\mu\partial_\nu \phi, \Box \phi}$, with spins $\ell=[0, 0, 1, 2, 0]$, implying a degeneracy pattern of $[1, 1, 3, 5, 1]$ for the lowest eigenvalues.
Except for $h\lesssim 5$, this expected degeneracy is observed over the plotted range.
Since the scaling dimensions $\Delta_{\Box \phi}$ and $\Delta_{\partial_\mu\partial_\nu \phi}$ are degenerate, numerically the correct pattern of spins appears as either $[0, 0, 1, 2, 0]$ or $[0, 0, 1, 0, 2]$.
As $N$ increases, the region along the critical line where the leading irreps are properly ordered becomes larger.
Given the good agreement with the expected irrep ordering, we now turn to the spectrum.

For given values of $(V_0,h)$, to compare the spectrum along the critical boundary  with CFT predictions, we must address two key issues.

\noindent \emph{a) What is the optimal  normalization $\alpha^{-1}$ of the Hamiltonian?}~\footnote{Often called the ``speed of light" in quantum phase transitions.} 

\noindent In the limit $N \rightarrow \infty$, the CFT lies exactly on the critical boundary—reached by tuning $g$ with infinite precision. 
However, at finite $N$:

\noindent \emph{b) does the spectrum of $H$ align more closely with the CFT for $g < g_c(h)$ or at $g = g_c(h)$? }

\noindent   As we will show in the next section, both questions share a common resolution.

\section{Tuning to the Lee-Yang CFT}


To guide our search for optimal tuning, we evaluate
\be
\chi^2_\text{min}(g)\equiv   \displaystyle\min_{\alpha}
 \sum_{i=1}^4 ( \delta E_i (\alpha, g) - \Delta_i)^2/\Delta_i  \, ,  \label{firstChi}
\ee
where 
$ \delta E_i \equiv  E_i -E_{\mathds{1}}$ are the energy gaps of the Hamiltonian \reef{LYham}.
For starters we keep $h$ and $V_0$ fixed, and only display the explicit dependence of the gaps on $\alpha$ and $g$.
The scaling dimensions are $\Delta_i \in [\Delta_\phi,\Delta_\phi{+}1,\Delta_\phi{+}2,\Delta_\phi{+}2]$ with $\Delta_\phi$ the  best estimate from the $\eps$-expansion.~\footnote{
The input scaling dimension varies within $\Delta_\phi \in [0.214, 0.216]$, but this variation is imperceptible at the y-axis scales of  Fig.~\ref{fig:three_plots}.}
Below we will  measure $\Delta_\phi$, but as a starting point, we first assess whether the current best estimate of $\Delta_\phi$ is compatible with realising the LY conformal spectrum.

In Fig.~\ref{fig:three_plots} (left) we show $\chi_\text{min}(g)$
for $N{ =} 14$ (dashed line with diamond markers) and  for $N {=} 16$ (solid line with triangle markers). 
Both lines are obtained with $(V_0, h) = (7,\, 29.75)$ -- details on $(V_0, h)$ tuning are given below.
Interestingly,  the minimum is attained for a value $g_\text{min}$ to the left of the critical value $g_c$ -- whose value is indicated inside the plot. 
Notably, the distance between $g_c$ and  $g_\text{min} $ decreases as $N$ increases from 14 to 16, consistent with  the expectation that $g_\text{min}$ tends to $g_c$ as $N\rightarrow \infty$.

\begin{figure*}[t!]
    \centering
  \includegraphics[width=0.485\textwidth]{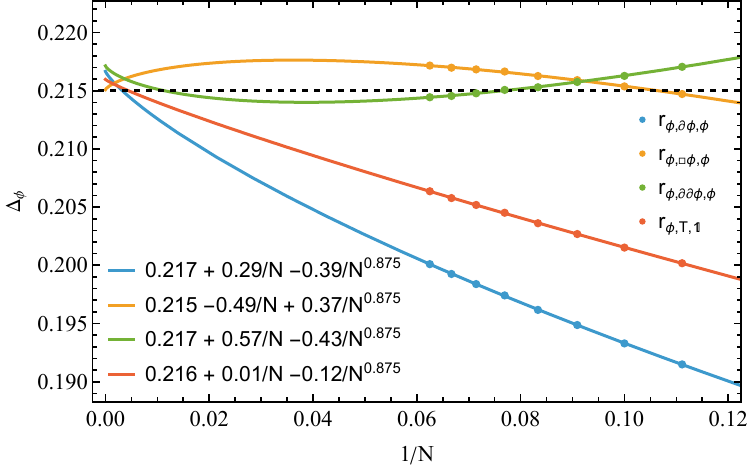}\hfill
  \includegraphics[width=0.47\textwidth]{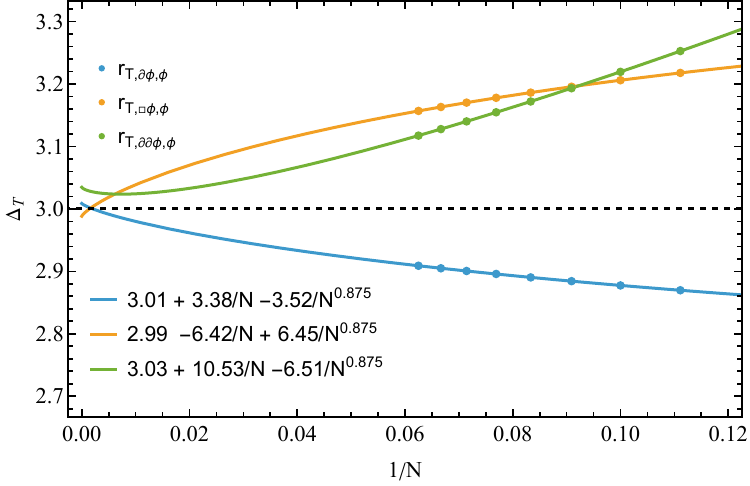}
   \caption{Large $N$ extrapolations of $\Delta_\phi$ and $\Delta_T$.  }  \label{fig:measurement}
\end{figure*}

\medskip 
The IR Effective Field theory provides an important guidance 
into how to tune the parameters of the UV model~\cite{Lao:2023zis,Lauchli:2025fii,Fardelli:2024qla}.
Indeed, to better understand the physical origin of the minimum of $\chi_\text{min}(g)$, we carry out a simple Effective Field Theory fit.
Consider the IR theory obtained by perturbing the LY CFT by the single relevant scalar, 
\vspace{-5pt}
\be
H_\text{IR} = H_\text{CFT}+ g_\phi V \quad \text{with}\quad V= \int_{S_2} d^2x \phi(x)  \, .  \label{Hir}
\ee
One can readily compute the first-order term in the CPT expansion of the LY CFT spectrum,
$
e_i(g_\phi) \equiv \Delta_i + g_\phi \langle i | V | i \rangle 
$. \footnote{The relevant formulas have been worked out in \cite{Lao:2023zis}, App. C. The corrections to all states in the $\phi$ multiplet are proportional to the (unknown) OPE coefficient $f^\phi_{\phi\phi}$, which in this work we absorb into $g_\phi$ via $g_\phi \to g_\phi / f^\phi_{\phi\phi}$. Note that this is only possible if considering only states from one conformal family.}
Next we evaluate
\be
\Sigma^2_\text{min}(g)\equiv   \displaystyle\min_{\{\alpha,\, g_\phi\}}
\sum_{i=1}^4 ( \delta E_i (\alpha, g) - e_i(g_\phi))^2/e_i (g_\phi) \, .  \label{secondChi}
\ee
In Fig.~\ref{fig:three_plots} (left) we show $\Sigma_\text{min}(g)$
for $N = 14$ (dashed line with round markers) and  for $N = 16$ (solid line with square markers). 
At this scale the curves look flat -- they have a tiny negligible slope.

In Fig.~\ref{fig:three_plots} (middle plot), we show the values of $g_\phi$ obtained from the minimization in \reef{secondChi}. Note that $g_\phi$ crosses zero at the minimum of $\chi_\text{min}(g)$, and $|g_\phi|$ increases with $N$ at fixed values of  $g-g_\text{min}$, as expected for a relevant perturbation—both providing an important consistency check.
For completeness we also show  the values of $\alpha$ as a function of $g$.

In summary, we identify $g_\text{min}$—the value that minimizes $\chi_\text{min}(g)$—as the point where the Hamiltonian spectrum~\reef{LYham} aligns most closely with that of the LY CFT, and this occurs strictly below $g_c$.

\medskip

Next we repeat the same exercise varying $h$. The result is shown in  the right plot of   Fig.~\ref{fig:three_plots}.
The value of $\text{min}_g \chi_\text{min}$ decreases as  $h$ increases. 
We eventually stop plotting for larger values of $h$, as  the  spins of the higher energy states not included in the fit \reef{firstChi}  deviate from expectations.~\footnote{After the degenerate states  $\Delta_{\Box \phi}$ or $\Delta_{\partial \partial \phi}$  we get $\{T_{\mu\nu},\partial\partial\partial\phi,\Box\partial\phi\}$. However as $h$ increases above the range of the plot, we observe the appearance of large spin $\ell>4$ states after these three states -- a feature not expected.}

Since, all the points in the curve of   Fig.~\ref{fig:three_plots} (right) correspond to   values of \reef{secondChi} with $g_\phi=0$,
the physical consequence of increasing $h$ is to tune  away the  coefficients of the leading irrelevant operator that can appear in the IR EFT -- more about it below.

In Fig.~\ref{fig:three_plots} (right) we also show  the same results for a number of different values of $V_0$ and $N$. 
What we observe is that for any of the values of $V_0$ explored, by sufficiently increasing  $h$ we can achieve comparably good $\chi_\text{min}$ values. Meanwhile the dependence on $N$ is very clear: for fixed $(V_0,h)$ the value of $\text{min}_g \chi_\text{min}$ decreases with increasing $N$.
While fine-tuning $V_0$ or including higher-order pseudo-potentials could offer further improvements, we do not pursue this direction here.

\section{Conformal spectrum}
\label{cs}

Having understood how to tune the model, we next turn our attention to the spectrum of the theory
for the best tuned value.

In Fig.~\ref{fig:multiplets} we show, with blue dots,  the energy gaps of the Hamiltonian \reef{LYham}. 
$\alpha$ has been  determined by minimizing $\chi(\alpha)_{N=16}$, see Fig.\ref{fig:three_plots}, with $\Delta_\phi=0.215$ -- variation of $\Delta_\phi$ within the best estimate range leads to similar results for   $\chi_\text{min}$.
The gaps shown on the plot are $
\delta E_i{=}[ 0.216, 1.278, 2.174, 2.199, 2.908, 3.095, 3.115] $
for spins $[0,1,0,2,3,2,1]$.
We measure the error of each energy gap  with respect to $\Delta_i$.
It is nontrivial that the value of the $\delta E_\phi$ gap remains this close to the input value of $\Delta_\phi$.
Overall, the spectral lines are strongly consistent with CFT expectations.

\subsection{Determining $\Delta_\phi$}

Given the strong agreement between the irreps and the spectrum with CFT expectations, we use the Fuzzy Sphere spectrum to extract $\Delta_\phi$. To this end, we repeat the procedure outlined above for a sequence of $N$ values. Specifically, we fix $\Delta_\phi = 0.215$ and determine $\{\alpha, g\}$ by minimising $\chi_\text{min}(g)$. For each $N$, the value of $g_\phi$ extracted from $\Sigma_\text{min}$ crosses zero precisely at the value of $g$ that minimises $\chi_\text{min}(g)$.

Having determined the optimal value of $g$ for each $N$, 
  we extract  $\Delta_\phi$ from the following ratios constructed from the  Hamiltonian spectrum:
 $r_{\phi,\partial\phi,\phi}$, $r_{\phi,\Box\phi,\phi},$ $r_{\phi,\partial\partial\phi,\phi}$ and $r_{\phi,T,\mathds{1}}$, where we define $r_{\mathcal{O}_1,\mathcal{O}_2,\mathcal{O}_3}\equiv (E_{\mathcal{O}_1}-E_\mathds{1})(\Delta_{\mathcal{O}_2}-\Delta_{\mathcal{O}_3})/(E_{\mathcal{O}_2}-E_{\mathcal{O}_3})$.
 If the spectrum matches that  of the LY  CFT, each of these ratios should converge to $\Delta_\phi$ as  $N\to \infty$.
 
 These ratios are independent of both  $\alpha$ and  the input value of $\Delta_\phi$. However, since the optimal value of $g$ was obtained by minimising   $\chi_\text{min}(g)$ \reef{firstChi} -- which does  on $\Delta_\phi$ -- consistency  requires that the ratios  converge   to the input value $\Delta_\phi=0.215$ as  $N\to \infty$.  
 We will subsequently vary $\Delta_\phi$, but let us first check whether this consistency requirement is met. 
 
In the left plot of Fig.~\ref{fig:measurement}, we present the results of this analysis, shown with round markers. While the ratios have not fully converged at this $y$-axis scale, their relatively modest variation is well accounted for by an Effective Field Theory fit, as we will show next.

In Fuzzy Sphere studies, $\alpha$ is often fixed by imposing that the energy-momentum tensor has dimension $\Delta_T = 3$. In contrast, in this work we determine $\alpha$, and the critical value of $g$,  by requiring that the IR EFT is tuned such that the coupling of the relevant operator, $g_\phi$, vanishes. This choice allows us to perform large-$N$ extrapolations within the framework of EFT.

Next we fit the ratios with the  functions $f(N)=a+b/N+c/N^{0.875}$.
The exponents follow from perturbation theory 
$E_i= \Delta_i + \alpha_i  N^{(3-\Delta_{\phi^\prime})/2} + \beta_i  N^{(3-\Delta_C)/2} $
and the value of the  leading irrelevant operators   $\Delta_{\phi^\prime}{\approx}5$ and $\Delta_{C}{\approx}4.75$.~\footnote{It is important to include the leading Lorentz breaking spin-4 operator $C$. More about it below. }
The extrapolated value of all the ratios is nicely  consistent with the  input $\Delta_\phi{=}0.215$, shown in the figure with a horizontal dashed line.

All fit parameters correspond to small corrections within the EFT. 
Notably, $r_{\phi,T,\mathds{1}}$ (red line) exhibits the smallest corrections. 
In the next section, we will use the ratios $r_{\mathcal{O},T,\mathds{1}}$, to extrapolate the values of several additional states.

In Fig.~\ref{fig:measurement}, right plot, we use an analogous procedure  to predict  the value of $\Delta_T$, which  must be three. This is a non-trivial crosscheck and our extrapolations are consistent with this expectation.

To test the discriminating power of our approach,
we repeated the analysis for several input values of $\Delta_\phi$, varying within a twenty-per-cent range. 
We find that the  Fuzzy Sphere data, along with  their EFT-based extrapolations, are compatible with this entire  range.
In order to improve  precision, this   exercise has to be refined by increasing $N$ and performing extrapolations of several states of the $\phi$ multiplet and crucially, other conformal families. 
Without this, the extrapolated Fuzzy Sphere spectrum remains compatible with a relatively broad range of $\Delta_\phi$ values.

\subsection{Effective Field Theory fit}

\begin{figure}[t]
  \centering
  \includegraphics[width=0.47\textwidth]{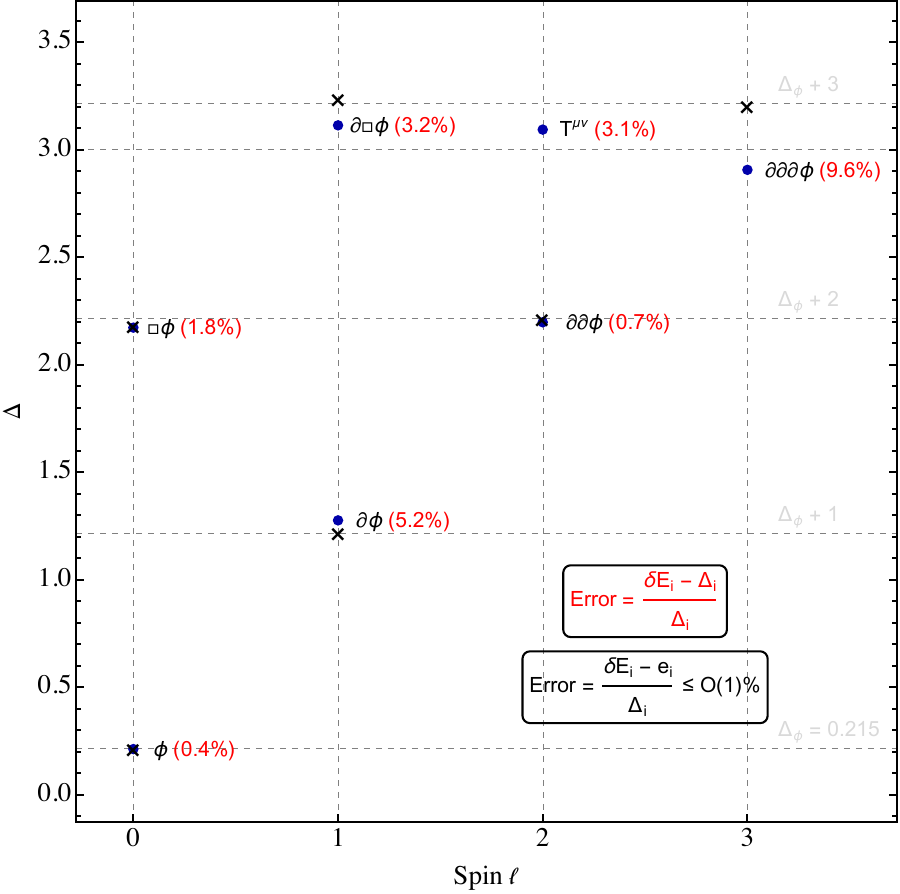}
  \caption{The lowest six states belonging to the $\phi$ conformal multiplet and the energy-momentum tensor.  Here $N=16$, and     the parameters   are $(V_0,h,g, \alpha){=}( 7,\, 29.75, \,  20.931, \, 0.079)$. 
   }
  \label{fig:multiplets}
\end{figure}

In Fig.~\ref{fig:multiplets} we have established  a good agreement between  the scaling dimensions of the relevant operators ($\Delta_i <3$)  and the CFT expectations. 
Nevertheless there are appreciable deviations in the higher energy states $\Delta_i \geq 3$. We will try to explain this remaining error via the following Effective Field Theory: 
\be
H_\text{IR} = H_\text{CFT}+ g_\phi  V_{\phi}+ g_{\phi^\prime}  V_{\phi^\prime}+  g_C V_C + \cdots \, ,  \label{Hirtwo}
\ee
where ellipses denote further higher-dimensional operators which are neglected,
and   the potentials $V_{\mathcal{O}}$ are the integrals of the local operators $\mathcal{O}(x)$ over the 2-sphere.
The scalar operator $\phi^\prime$ is $\phi^3$ in the GL description~\reef{GL};~\footnote{In the 2D LY EFT the leading irrelevant operator would be $T\bar T$, while $\phi^3$ would correspond to the first non-total-derivative Virasoro descendant (at level 4) of $\phi$~\cite{Xu:2022mmw,Gliozzi:2014jsa}.}
 the spin-4 operator $C_{\mu\nu\rho\sigma}$ is contracted with four unit vectors on the sphere  to form a scalar, and is the leading Lorentz-breaking operator in the EFT --  an analogous effect has been observed in the Ising EFT~\cite{Lauchli:2025fii}.

Interestingly $\phi^\prime$ does not correct all the states in the $\phi$-multiplet with the same sign -- for a given $g_{\phi'}$ some are shifted upwards while others downwards.  
It turns out that the pattern of the corrections is such that  $\phi^\prime$ alone is not enough to explain the red errors  in Fig.~\ref{fig:multiplets}.

We are thus forced to include the spin-4 operator $C$, whose scaling dimension $\Delta_C$ is  comparable  to $\Delta_{\phi^\prime}$.  
Next we perform a fit by minimising the following quantity:
\be
\Xi_\text{min}^2  (g) \equiv  \displaystyle\min_{\{\alpha,\, \vec{g}\}} \sum_{i=1}^6   ( \delta E_i (\alpha, g) - e^\text{CPT}_i(\vec{g}))^2/e^\text{CPT}_i (\vec{g}) \, ,  \label{thirdChi}
\ee
where $i$ runs over the  first six states in the $\phi$-conformal family, $\vec{g}=(g_{\phi},g_{\phi^\prime},g_c)$, and $e_i^{\text{CPT}}(\vec{g}) \equiv \Delta_i + g_\phi \bra{i}V_\phi \ket{i} + g_{\phi^\prime} \bra{i}V_{\phi^{\prime}}\ket{i}+g_C\bra{i}V_C\ket{i}$.
The results are shown in Fig.~\ref{fig:multiplets},  where  black-cross markers denote   $\delta E_i {-} (e_i^\text{CPT}{-}\Delta_i)$
are evaluated  at the point where $\Xi_\text{min}(g)$ selects $g_\phi \approx 0$. 
This occurs at a value of $g$ slightly away from $g_\text{min}$ obtained by minimizing $\chi_\text{min}(g)$, which can be attributed to the presence of the irrelevant operators in the fit.

Both the $g_{\phi^\prime}$ and $g_C$ corrections are individually nicely perturbative. 
The errors on the scaling dimensions of the operators $\{\phi, \partial \phi, \Box \phi, \partial\partial \phi, \partial\partial  \partial \phi, \partial \Box\phi \}$ decrease significantly from the ones reported in red in Fig.~\ref{fig:multiplets}, and are given by $\{0.12, 0.55, 1.36, 0.17, 0.19,  0.78 \}\%$,  i.e. sub-percent  for most of  them. We do not include the $T_{\mu\nu}$ state because, as it is in a different conformal multiplet, we do not know its correction $e_T^{\text{CPT}}(\vec{g})$ without knowing the OPE coefficients $f^{\phi}_{TT},f^{\phi'}_{TT}$ and $f^{C}_{TT}$. It would be interesting to measure them  using the method recently proposed in Ref.~\cite{Lauchli:2025fii}, 
which would  only require knowing $f_{\phi\phi}^\phi$ as an input.

\subsection{Large $N$ Extrapolation of the  spectrum}

In the previous section, we showed that the deviations of the low-energy spectrum in the $\phi$-multiplet from the conformal predictions could be accurately captured by an Effective Field Theory (EFT) fit.
In this section, we extend the analysis to many higher-energy states. 
Rather than performing another EFT fit, we instead extrapolate to the $N \to \infty$ limit.
To do so, we compute the ratio $r_{\mathcal{O},T,\mathds{1}}$ for each state $|\mathcal{O}\rangle$ and for each volume $1/N$ -- with  $N=9,10,11,{\dots},16$.

We emphasize that, for each $N$, we have tuned the coupling $g$ to ensure that the EFT coupling of the relevant operator $\phi$ vanishes, i.e. $g_\phi \approx 0$ -- minimizing $\chi_\text{min}(g)$ with $\Delta_\phi=0.215$.
We then fit the energy levels using the EFT based function $f(N) = a + b/N + c/N^{0.875}$, which includes tree-level corrections from the leading irrelevant primary operators $\phi^3$ and $C_{\mu\nu\rho\sigma}$, of spin zero and four, respectively.

\begin{figure}[t]
  \centering
  \includegraphics[width=0.47\textwidth]{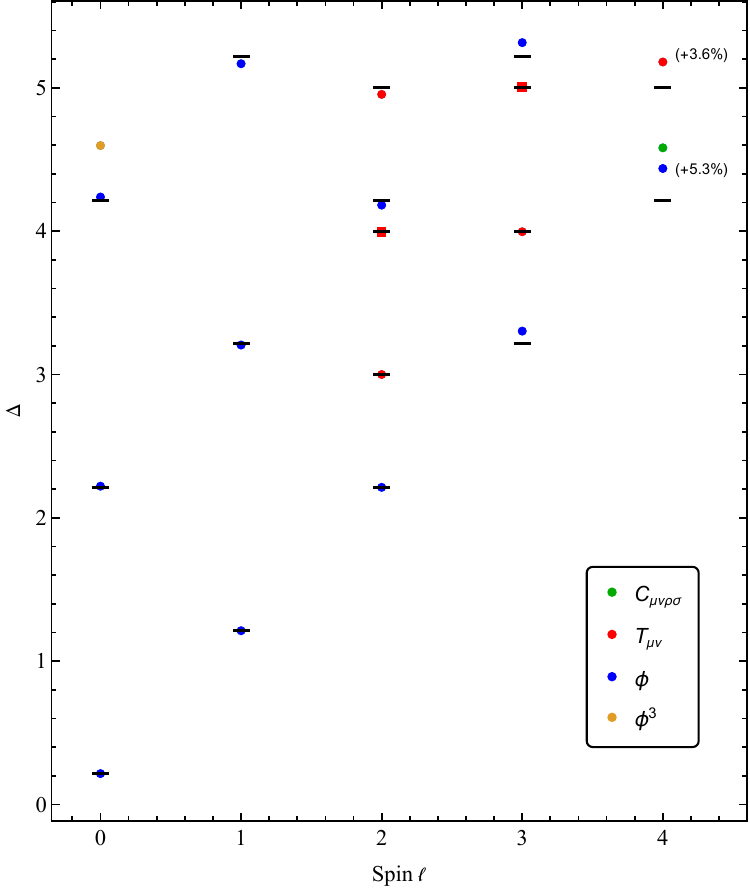}
  \caption{Low-lying  multiplets in the 3D Lee-Yang CFT.}
  \label{fig:extramultiplets}
\end{figure}

The outcome of this analysis is shown in Fig.~\reef{fig:extramultiplets}, which displays the first nineteen states with spin $\ell \leq 4$.
Horizontal lines indicate the extrapolated values $\Delta_\phi + n$ and $\Delta_T + n$  for each descendant level $n$, while  markers represent the extrapolated energy levels (round for parity even and square for parity odd).
The deviation between markers and lines quantifies the departure from conformality.

The outcome of this analysis is self-consistent: as indicated by the orange and green markers, the states $|\phi^3\rangle$ and $|C\rangle$ have a predicted energy of $4.6$ and $4.58$ respectively, corresponding to  deviations of less than $10\%$ and $5\%$, respectively, from the values used in the extrapolation functions $f(N)$.
The largest errors -- associated with the spin-4 descendants -- are indicated in the plot. All other states have errors below $3\%$, with most below $1\%$!

\section{Conclusions}

A  number of works have employed the Fuzzy Sphere approach to study several other observables of the 3D Ising CFT~\cite{Hu:2023xak,Han:2023yyb,Zhou:2023fqu,Hu:2023ghk,Zhou:2024dbt,Cuomo:2024psk,Dedushenko:2024nwi,Hu:2024pen}, 
as well as unitary 3D CFTs other than the Ising~\cite{Zhou:2023qfi,Chen:2023xjc,Han:2023lky,Chen:2024jxe,Zhou:2024zud,TalkLauchli,Voinea:2024ryq,Yang:2025wqn}.

In this work, we have opened the possibility of using the Fuzzy Sphere approach to study the LY CFT, paving the way for several avenues of further investigation.

For instance, it would be interesting to generalise this approach to other  non-unitary 3D CFTs. 
GL description of non-unitary CFTs has been a topic of recent attention~\cite{Klebanov:2022syt,Katsevich:2024jgq}.
A number of these descriptions are valid in 3D, and a picture is emerging which points to the existence of several  non-unitary  3D CFT and a rich structure of RG flows connecting them and unitary CFTs such as the Ising. The 3D fuzzy sphere approach can be very valuable in solidifying this picture.

As for the LY CFT, there are a number of interesting improvements ahead. 
These include refining the determination of $\Delta_\phi$,  conducting a more detailed study of the Lee-Yang EFT near the critical point,  extracting  OPE coefficients~\cite{Hu:2023xak} -- see also \cite{Lauchli:2025fii} -- , constructing the conformal generators \cite{Fardelli:2024qla,Fan:2024vcz},  and achieving more rigorous large $N$ extrapolations. We look forward to exploring these and other aspects of the fuzzy sphere regulator.

\bigskip 
\noindent \emph{Note added.} While finalising this manuscript, we became aware that \cite{ArguelloCruz:2025zuq,Fan:2025bhc} were independently investigating the 3D Lee-Yang CFT on the Fuzzy Sphere.

 \section*{Acknowledgements}
 We thank Nicola Dondi,   Giulia Fardelli, Aditya Hebbar, James Ingoldby, Slava Rychkov and Ling-Xiao Xu for valuable discussions. 
 JEM  is supported by the European Research Council, grant agreement n. 101039756.



\small

\bibliographystyle{utphys}

\bibliography{biblio}



\appendix
\onecolumngrid

\section{ 
Reproducing von~Gehlen's~\cite{vonGehlen:1991zlm}: 2D Lee--Yang quantum  critical point}
\label{vG}

We provide a simple  \textit{Mathematica} code to reproduce the results of von~Gehlen~\cite{vonGehlen:1991zlm}, see also~\cite{Castro-Alvaredo:2009xex},  which we found serves as a model example for the more complex system studied here.
The spectrum of the 2D Ising model can be retrieved from the spin-chain Hamiltonian \verb|H[\[Lambda]_]|
 defined here:
\begin{verbatim}
Nv = 15; (*number vertices*)
edges = Join[Table[{i,i+1},{i,1,Nv-1}],{{Nv,1}}];
\[Sigma]x = SparseArray[{{1,2}->1.,{2,1}->1.}];
\[Sigma]z = SparseArray[{{1,1}->1.,{2,2}->-1.}];
id2 = SparseArray[{{1,1}->1.,{2,2}->1.}];

Hx = Plus@@(F[\[Sigma]x,#]&/@Range[1,Nv]);
F[mat_, index_] := 
  KroneckerProduct@@(Table[id2,{i,1,index-1}]~Join~{mat}~Join~Table[id2, {i,index+1,Nv}]);
Hzz = SparseArray[Plus@@(Dot[(F[\[Sigma]z,#]&/@ 
           Range[1,Nv])[[#[[1]]]],(F[\[Sigma]z,#]&/@ 
           Range[1,Nv])[[#[[2]]]]]&/@edges)];

H[\[Lambda]_] := -1/2(Hzz+\[Lambda] Hx);
eigs[\[Lambda]_] := Eigenvalues[H[\[Lambda]]-1000IdentityMatrix[2^Nv,SparseArray],50]+1000;

Drop[Nv/(2\[Pi])eigs[1]-Nv/(2\[Pi])eigs[1][[1]],1][[1;;10]]
\end{verbatim}
The output of the last line is to be compared with the  ten lowest energy  states   of the 2D Ising CFT.
This spin-chain is integrable.
As \texttt{Nv} is increased further we obtain nice convergence to the 2D Ising CFT. 
This code  can be improved in a number of simple ways, nevertheless it can serve as a helpful starting point. 
In order to get the LY fixed point of Ref.~\cite{vonGehlen:1991zlm} one needs to deform the previous Hamiltonian by \verb|H[\[Lambda]_]| $\rightarrow $  \verb|H[\[Lambda]_]+i h Hz| where 
\begin{verbatim}
Hz = Plus@@(F[\[Sigma]z,#]&/@Range[1,Nv]);
\end{verbatim}

We have also studied 3D spin lattices generalisations of this Hamiltonian employing the Icosahedron~\cite{Lao:2023zis} and Dodecahedron platonic solids. These lattices have Icosahedral symmetry, the  largest finite subgroup of $SO(3)$. 
This can be achieved by changing the variable $\texttt{Nv}$ and \texttt{edges} (second line of the code above) by 
\begin{verbatim}
Nv = 12 (*number vertices*)
edges = (List@@#&)/@EdgeList[GraphData["IcosahedralGraph"]];
\end{verbatim}
or \texttt{Nv = 20} together with  \texttt{edges=(List@@\#)\&/@EdgeList[GraphData["DodecahedralGraph"]];}.
These relatively small 3D lattices (\texttt{Nv} = 12 and 20) can be efficiently handled in \textit{Mathematica}.
This code can be used to reproduce the results of~\cite{Lao:2023zis}. As shown there, when combined with Conformal Perturbation Theory, these coarse-grained lattices encode a wealth of information.

\twocolumngrid

\end{document}